\documentclass{Interspeech2024}
\pdfoutput=1

\interspeechcameraready

\usepackage{multirow}
\usepackage{graphicx}
\usepackage{booktabs} 
\title{Towards Realistic Emotional Voice Conversion  using \\ Controllable Emotional Intensity}

\name[affiliation={1,2}]{Tianhua}{Qi}
\name[affiliation={1,2}]{Shiyan}{Wang}
\name[affiliation={1,2}]{Cheng}{Lu}
\name[affiliation={1}]{Yan}{Zhao}
\name[affiliation={1,2}]{Yuan}{Zong}
\name[affiliation={1,2}]{Wenming}{Zheng$^*$}

\address{
  $^1$Key Laboratory of Child Development and Learning Science (Southeast University),\\ Ministry of Education, Nanjing 210096, China \\
  $^2$School of Biological Science and Medical Engineering, Southeast University, China 
}
\email{qitianhua@seu.edu.cn, xhzongyuan@seu.edu.cn, wenming\_zheng$^{*}$@seu.edu.cn}


\begin{document}

\maketitle

\begin{abstract}
Realistic emotional voice conversion (EVC) aims to enhance emotional diversity of converted audios, making the synthesized voices more authentic and natural. To this end, we propose Emotional Intensity-aware Network (EINet), dynamically adjusting intonation and rhythm by incorporating controllable emotional intensity. 
To better capture nuances in emotional intensity, we go beyond mere distance measurements among acoustic features. Instead, an emotion evaluator is utilized to precisely quantify speaker's emotional state. By employing an intensity mapper, intensity pseudo-labels are obtained to bridge the gap between emotional speech intensity modeling and runtime conversion.
To ensure high speech quality while retaining controllability, an emotion renderer is used for combining linguistic features smoothly with manipulated emotional features at frame level. Furthermore, we employ a duration predictor to facilitate adaptive prediction of rhythm changes condition on specifying intensity value.
Experimental results show EINet's superior performance in naturalness and diversity of emotional expression compared to state-of-the-art EVC methods. 
\end{abstract}

\section{Introduction}

\textit{``Everything I read I try to figure out: what it really means, what it’s really saying.''}

\textit{\qquad\qquad\qquad\qquad\qquad\qquad\quad\quad\quad\---Richard P. Feynman}

Emotional voice conversion (EVC) endeavors to transform the state of a spoken utterance from one emotion to another, while preserving the linguistic content and speaker identity~\cite{yang2022overview}. It holds the promise of fostering more profound emotional communication between individuals~\cite{zhou2020converting}, elevating user experience in human-machine interactions~\cite{erol2019toward}, as well as creating a more immersive and resonant virtual experience~\cite{byeon2022voice}.

Current EVC systems are predominantly constructed based on autoencoder~\cite{kim2020emotional, zhou2021seen, lu2022one, zhuemotional2023} especially for sequence-to-sequence (seq2seq)~\cite{chen2022speaker, qi2024pavits} frameworks, with significant strides in speech quality.
However, the converted audio lacks emotional diversity, which is a critical aspect for achieving realistic speech synthesis.
Therefore, incorporating an intensity control module into typical EVC framework has become a primary research focus to facilitate manipulation of emotional expression, consequently addressing one-to-many problem in a controllable manner.
\let\thefootnote\relax\footnotetext{Speech samples are available at \url{https://jeremychee4.github.io/EINet4EVC/}.}

For instance, Emovox~\cite{zhou2022emotion} is constructed based on Seq2seq-EVC~\cite{zhou2021limited}, leveraging the relative attribute ranking (RAR)~\cite{parikh2011relative} metric to measure relative difference among acoustic features such as pitch frequency and frame energy, between emotional and non-emotional speech samples. Additionally, intensity pseudo-labels are generated to address the absence of explicit annotations in emotional corpora~\cite{zhou2022emotional}. 
Similarly, AINN~\cite{chen2023attention} is built upon EmotionalStarGAN~\cite{rizos2020stargan}, incorporating contrastive learning to construct positive and negative pairs. The calculation of intensity pseudo-labels is also employed to control emotion transformation by explicitly specifying an intuitive intensity value. 

Despite the great success achieved by intensity control approaches in EVC, the converted vocal expressiveness still falls short of meeting human perceptual expectations, particularly in terms of naturalness and diversity. This deficiency can be attributed to the prevalent utilization of intensity modeling methods that solely rely on measuring the differences in acoustic features~\cite{zhou2022emotion, chen2023attention, lei2022msemotts, zhou2022speech}. This dependency overlooks inherent emotional fluctuations of speaker, leading to a mismatch between emotional intensity modeling and run-time conversion, which poses a substantial obstacle to synthesize authentic voices. 

The dimensional representation method allows for a more accurate portrayal of the distinctions between emotional states, drawing inspiration from the circumplex theory~\cite{russell1980circumplex}.
As proposed by ~\cite{reisenzein1994pleasure}, within the 2-dimensional VA-space formed by valence and arousal, the wedge area formed by these dimensions can be utilized to gauge emotional intensity values.
Consequently, incorporating valence-arousal-dominance (VAD) values into the emotional intensity control module offers a promising approach for achieving precise generation of emotional intensity pseudo-labels, along with nuanced control over emotional expression.

Based on above considerations, we propose the Emotional Intensity-aware Network (EINet) that leverages controllable emotional intensity to enhance naturalness and diversity of converted audios, ultimately advancing emotion conversion towards more realistic synthesis.
In contrast to solely measuring acoustic features, we focus on the distance between emotional features to construct pseudo-labels for emotional intensity, which effectively addresses the mismatch between emotional intensity modeling and run-time conversion in EVC.
To discern nuances in emotional expression at utterance level, the emotion evaluator is employed to anticipate the  valence-arousal-dominance (VAD) values behind the speech.
Distances between VAD values are further assessed by intensity mapper to obtain pseudo-labels that better align with human-perceived emotional intensity, which highly contributes to enhancing the emodiversity of converted audios.
To ensure that speech quality is not compromised by intensity control, the emotion renderer is utilized to integrate  linguistic features and manipulated emotional features at frame level.
Additionally, we use duration predictor to modify speech duration, adaptively forecasting rhythmic alterations based on emotional intensity values.  

\begin{figure*}[ht]
	\centering
	\includegraphics[width=2.1\columnwidth]{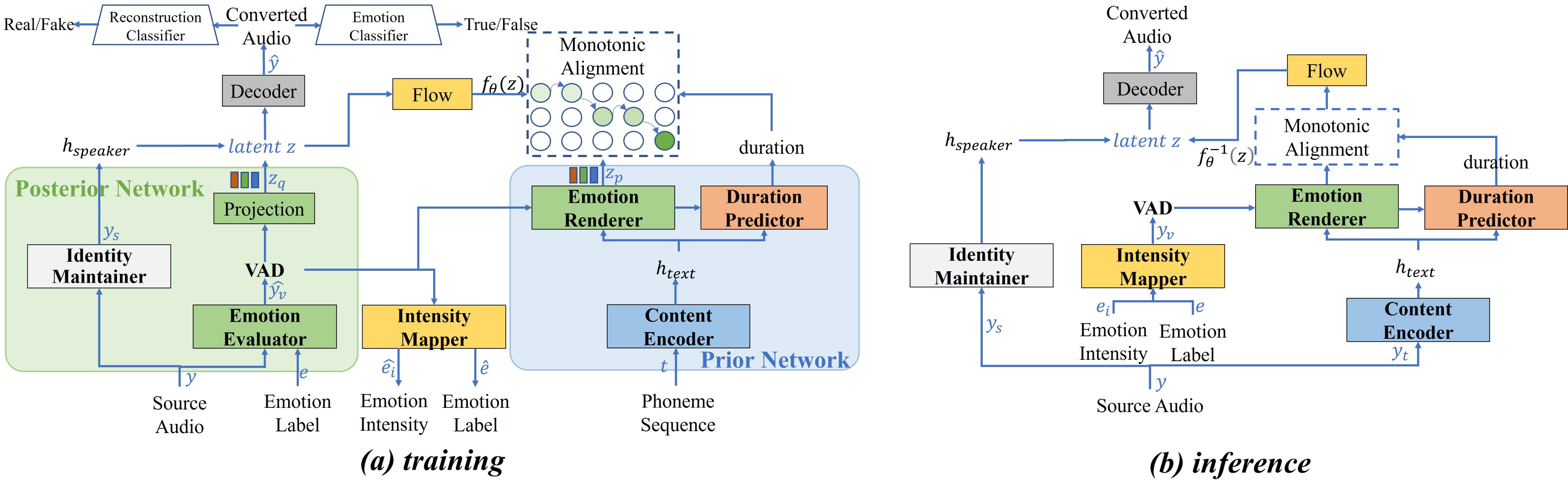}
	\caption{Diagram of our proposed EINet, depicting the training procedure(a) and inference procedure(b).}
	\label{Fig1}
\end{figure*} 

\section{Proposed Method}
As depicted in Figure 1, EINet is constructed based on conditional variational autoencoder (CVAE), consisting of a posterior network, an intensity mapper, a prior network, and a decoder.

The posterior network (PosNet) captures the inherent emotional states, i.e., valence-arousal-dominance (VAD) values, denoted as $\mathcal\ y_v$, from the source audio $\mathcal\ y$ given specific emotion category $\mathcal\ e$, serving as a condition factor for generating posterior distribution $\mathcal\ q\left(z_q \mid c_q\right)$. Besides, speaker characteristic $\mathcal\ h_{speaker}$ is extracted to alleviate the issue of identity loss.
\begin{equation}
	\setlength\abovedisplayskip{4pt}\setlength\belowdisplayskip{4pt}
	\begin{aligned}
		z_q=PosNet\left(c_q\right) \sim q\left(z_q \mid c_q\right) 
	\end{aligned}
	\label{eq1}
\end{equation}
where $\mathcal\ c_q$ including source audio $\mathcal\ y$ and emotion category $\mathcal\ e$.

The intensity mapper (IM) constructs intensity pseudo-labels $\mathcal\ \hat{e_i}$ based on inherent emotional states $\mathcal\ y_v$ during training, and generates corresponding VAD values $\mathcal\ \hat{y_v}$ given target emotion category $\mathcal\ e$ with specified intensity $\mathcal\ e_i$ during inference.
\begin{equation}
	\setlength\abovedisplayskip{4pt}\setlength\belowdisplayskip{4pt}
	\left\{\begin{aligned}\label{eq2}
		\hat{e} , \hat{e}_i & =I M\left(y_v\right) \\
		\hat{y}_v & =I M^{-1}\left(e, e_i\right)
	\end{aligned}\right.
\end{equation}

The prior network (PriorNet) predicts prior distribution $\mathcal\ p\left(z_p \mid c_p\right)$ based on linguistic content $\mathcal\ y_t$ and VAD values $\mathcal\ y_v$ containing intensity information.
\begin{equation}
	\setlength\abovedisplayskip{4pt}\setlength\belowdisplayskip{4pt}
	\begin{aligned}
		z_p=PriorNet\left(c_p\right) \sim p\left(z_p \mid c_p\right) 
	\end{aligned}
	\label{eq3}
\end{equation}
where $\mathcal\ c_p$ including linguistic content $\mathcal\ y_t$ and emotional descriptor $\mathcal\ y_v$.

The decoder reconstructs waveform according to latent representation $\mathcal\ z$, where $\mathcal\ z$ is derived from $\mathcal\ z_q$ during training and $\mathcal\ z_p$ during inference, both reinforced with identity information $\mathcal\ h_{speaker}$. Additionally, adversarial learning is employed to enhance naturalness progressively in content and emotion aspects.
\begin{equation}
	\setlength\abovedisplayskip{4pt}\setlength\belowdisplayskip{4pt}
	\begin{aligned}
		\hat{y}=Decoder\left(z\right) \sim p\left(y\mid z\right)
	\end{aligned}
	\label{eq4}
\end{equation}
\subsection{Posterior network}
Given $\mathcal\ c_q$ including source audio $\mathcal\ y$ and its emotion category $\mathcal\ e$, the posterior network provides the posterior distribution $\mathcal\ q\left(z_q \mid c_q\right)$ for CVAE. The emotion evaluator is utilized to extract VAD values at utterance level, which are further transformed to fine-grained emotional acoustic features (a fine-grained normal distribution with mean $\mathcal\ \mu_\theta$ and variance $\mathcal\ \sigma_\theta$ generated by a normalizing flow $\mathcal\ f_\theta$). Speaker characteristics $\mathcal\ h_{speaker}$ is also extracted by identity maintainer.
\begin{equation}
	\setlength\abovedisplayskip{4pt}\setlength\belowdisplayskip{4pt}
	\begin{aligned}
	q\left(z_q \mid c_q\right) = N\left(f_\theta\left(z_q\right) ; \mu_\theta\left(c_q\right) ; \sigma_\theta\left(c_q\right)\right) \\ 
	\end{aligned}
	\label{eq5}
\end{equation}

\textit{Emotion evaluator}: Since speech emotion is inherently supra-segmental, it is difficult to learn emotional latent representation and quantify emotional states in a proper manner. To tackle this, a specific speech emotion recognition (SER) model~\cite{wagner2023dawn} based on circumplex theory is introduced to predict the valence-arousal-dominance values $\mathcal\ \hat{y_t}$ for each utterance, which to assess the pleasantness, activation, and influentiality of speaker's internal states. Utilizing this prior knowledge, emotionally sensitive acoustic features can be extracted by two $1\times1$  convolutional layers with a Wavenet residual block, and expand to frame-level by a linear projection layer.

\textit{Identity maintainer}: Considering that controllable EVC has more manipulation over the synthesis of acoustic features, which makes it highly susceptible to speaker identity loss. Recognizing the importance of fundamental frequency ($\mathcal\ F_0$) with voicing flag ($\mathcal\ v$) in modeling speaker characteristics~\cite{busso2009analysis}, especially for intonation, we enhance the F0 predictor of ~\cite{zhang2022visinger} by incorporating a $1\times1$  convolutional layer and a linear layer to address this issue.
\begin{equation}
	\setlength\abovedisplayskip{4pt}\setlength\belowdisplayskip{4pt}
	\begin{aligned}
		L_{F_0}=||\log F_0-\log \hat{F}_0||_2+||v-\hat{v}||_2
	\end{aligned}
	\label{eq6}
\end{equation}

\subsection{Intensity mapper}
Different from solely quantifying differences between acoustic features using distance measurement methods, we utilize the intensity mapper to implicitly generate a distribution of intensity pseudo-labels based on emotion category $\mathcal\ e$ and VAD values $\mathcal\ y_v$, facilitating supervised training.
In order to establish a mapping relationship between intensity distribution and VAD values, it is constructed based on reversible normalizing flow~\cite{kobyzev2020normalizing}.

\textit{Intensity label construction}: During training, the intensity mapper utilizes VAD values extracted by the emotion evaluator to calculate pseudo-labels $\mathcal\ \hat{e_i} \in (0,1) $ for each sample and predict the emotional category $\mathcal\ \hat{e}$. 
\begin{equation}
	\setlength\abovedisplayskip{4pt}\setlength\belowdisplayskip{4pt}
	\begin{aligned}
	p\left(\hat{e}, e_i \mid y_v\right) = N\left(f_\theta\left(\hat{e}\right), f_\theta\left(e_i\right); \mu_\theta\left(y_v\right) ; \sigma_\theta\left(y_v\right)\right) \\ 
	\end{aligned}
	\label{eq7}
\end{equation}

To ensure precise mapping and fine-grained control, cross-entropy loss and feature mapping loss are both used to evaluate the accuracy of predictions at both categorical and feature levels. 
Furthermore, we introduce two coefficients to balance these losses during the early-to-mid and mid-to-late stages.
\begin{equation}
	\setlength\abovedisplayskip{2pt}\setlength\belowdisplayskip{2pt}
	\resizebox{.8\hsize}{!}{$
	\begin{aligned}
		L_{IM} &=\gamma \mathbb{E}_y\left[-\sum_{k=1}^K p_k \log \left(q_k\right)\right] \\
		&+ \beta \mathbb{E}_{(y, z_\text{IM})}\left[\sum_{l=1}^L \frac{1}{N_l}||D^l(y)-D^l(G(z_\text{IM}))||_1\right]
	\end{aligned}
	$}
	\label{eq8}
\end{equation}
where $\mathcal\ K$ represents the total number of emotional categories, $\mathcal\ p$ indicates the emotion label, $\mathcal\ q$ represents the classification distribution,$\mathcal\ L$ denotes the total number of layers of  discriminator, $\mathcal\ D^l$ indicates the $\mathcal\ l$-th layer feature map of the discriminator, with $\mathcal\ N_l$ denoting the number of features.

\textit{Emotional intensity control}: During inference, by specifying target emotion category $\mathcal\ {e}$ and intensity value$\mathcal\ e_i \in (0,1) $, 
the intensity mapper anticipates VAD values $\hat{y}_v$, enabling the direct modulation of emotional expression without the need for any reference audio.
\begin{equation}
	\setlength\abovedisplayskip{4pt}\setlength\belowdisplayskip{4pt}
	\hat{y}_v=f_{\theta}^{-1}\left(e, e_i\right) \\ 
	\label{eq9}
\end{equation}

\subsection{Prior network}
The prior network provides prior distribution $\mathcal\ p\left(z_p \mid c_p\right)$ for CVAE based on linguistic content $\mathcal\ y_t$ and emotion descriptors $\mathcal\ y_v$. The content encoder takes phoneme sequences as input to extract detailed linguistic feature $\mathcal\ h_{text}$ at first. 
To attain accurate control over emotions, the emotion renderer generates frame-level acoustic features based on VAD values. 
The duration predictor incorporates emotional and textual features to analyze the correlation among emotional intensity, linguistic sequence, and speech duration, which allows for the prediction of varying durations based on emotional intensity, ultimately enriching the overall emotional diversity.
\begin{equation}
	\setlength\abovedisplayskip{4pt}\setlength\belowdisplayskip{4pt}
	\resizebox{.89\hsize}{!}{$p\left(z_p \mid c_p\right) = N\left(f_\theta\left(z_p\right) ; \mu_\theta\left(c_p\right) ; \sigma_\theta\left(c_p\right)\right)\left|\operatorname{det} \frac{\partial f_\theta\left(z_p\right)}{\partial z_p}\right| \\$} 
	\label{eq10}
\end{equation}

\textit{Content encoder}: To avoid mispronunciations as well as skipping-words that significantly influence human perception, the content encoder plays a crucial role in extracting linguistic features from phoneme sequences, which ensures the preservation of textual content particularly in emotion conversion with intensity control. It comprises a fully connected layer, a Feed-Forward Transformer (FFT) block with a linear projection layer. 

\textit{Emotion renderer}: In order to seamlessly integrate emotional states with linguistic content, the emotion renderer expands generalized VAD values to nuanced emotional acoustic features. It involves a $1 \times 1$ dialted convolution layers, a Wavenet residual blocks, and a linear projection layer.

\textit{Duration predictor}: Considering that diverse emotional intensities can result in distinct voicing durations and pause locations, even when the textual content is the same, we integrate emotional feature and linguistic feature into the duration predictor, aiming to calculate the logarithm of duration at phoneme level, thereby substantially improving the rhythmic modeling capacity of emotional speech. It is constructed using five $1 \times 1$ convolution layers, two $1 \times 1$ dialted convolution layers and a linear projection layer.
\subsection{Final loss}
By combining CVAE with adversarial training, we formulate the overall loss function as follows:
\begin{equation}
	\setlength\abovedisplayskip{3pt}\setlength\belowdisplayskip{3pt}
	L=L_{cls }+L_{fm }+L_{adv }(G)+L_{F_0 }+L_{dur }+L_{IM }
	\label{eq11}
\end{equation}
\begin{equation}
	\setlength\abovedisplayskip{3pt}\setlength\belowdisplayskip{3pt}
	L(D)=L_{adv }(D)
	\label{eq12}
\end{equation}
where $L_{cls }$ minimizes the L1 distance between generated and target spectrogram, $L_{fm }$ minimizes the L1 distance between feature maps extracted from intermediate layers in each discriminator for a better training stability, $L_{adv }(G)$ and $L_{adv }(D)$ represent the adversarial loss for the Generator and Discriminator respectively, $L_{dur}$ minimizes the L2 distance between predicted duration and ground truth which is obtained through estimated alignment.

\section{Experiments}

\subsection{Experimental setup}
\textbf{Dataset.} We conduct emotion conversion using a Mandarin corpus within the Emotional Speech Dataset (ESD)~\cite{zhou2022emotional} from neutral to angry, happy, sad, and surprise, denoted as \textit{Neu-Ang}, \textit{Neu-Hap}, \textit{Neu-Sad}, \textit{Neu-Sur} respectively. The average durations for utterances in each emotional category are 3.23s, 2.68s, 2.84s, 4.04s, and 3.32s, respectively.  \\
\textbf{Data preparation.} For each conversion pair, the corpus is partitioned into a training set (300 samples, approximately 16 minutes), a validation set (30 samples), and a test set (20 samples). In our experiments, we employ  speech data represented by an 80-dimensional Mel-spectrogram extracted from audio recorded at a sampling rate of 16kHz. \\
\textbf{Implementation details.} Our proposed architecture is built upon VITS~\cite{kim2021conditional}, utilizing the AdamW optimizer with an initial learning rate of 2e-4, and a learning rate decay of 0.999875. Dropout probability is set at 0.1. The $\gamma$ coefficient starts at 1.00 and decreases by 0.01 every 5 epochs until it reaches 0.30. The $\beta$ coefficient is defined as $1-\gamma$. \\
\textbf{Models for comparison.} We train the following models to assess the effectiveness of proposed method. 
\begin{itemize}
	\item Seq2seq-EVC~\cite{zhou2021limited} (\textit{baseline}): a seq2seq-based EVC model supports basic emotion conversion but lacks controllability. 
	\item Emovox~\cite{zhou2022emotion} (\textit{baseline}): a seq2seq-based EVC model using RAR to calculate the distance between acoustic features, intensity pseudo-labels are obtained to facilitate control. 
	\item VITS-EVC~\cite{kim2021conditional} (\textit{baseline}): a EVC model constructed by original VITS, only supports basic emotion conversion.
	\item EINet (\textit{proposed}): the proposed model utilizing intensity mapper to compute the distance among VAD values, emotional intensity pseudo-labels are obtained to support control.
\end{itemize}

\subsection{Model performance}
\begin{table*}[]
	\centering
	\renewcommand{\arraystretch}{1.1}
	\setlength{\abovecaptionskip}{0.2cm}
	\setlength{\belowcaptionskip}{-0.4cm}
	\caption{Quantitative comparisons of converted speech with previous methods. The * denotes methods pretrained on VCTK corpus~\cite{veaux2017cstr}.}
	\label{tab:my-table}
	\resizebox{2\columnwidth}{!}{%
		\begin{tabular}{c|c|cccc|cc}
			\hline
			\multirow{2}{*}{EVC Model} & \multirow{2}{*}{Source of Intensity Pseudo-Label} & \multicolumn{4}{c|}{Objective Evaluation}                              & \multicolumn{2}{c}{Subjective Evaluation}                               \\ \cline{3-8} 
			&                      & MCD $\downarrow$ & $\text{RMSE}_{F_0 }$ $\downarrow$ & DDUR $\downarrow$               & $\text{ACC}_{cls }$ $\uparrow$  & Naturalness $\uparrow$               & Similarity $\uparrow$              \\ \hline
			Seq2seq-EVC*            & None                      & 4.22                        & 45.88             & 0.39              & 98.85\%                & 4.04±0.16                 & 67.97\%                           \\
			Emovox*                 & Acoustic features          & 4.23           & 47.13               & 0.36                  & 98.79\%                & 3.93±0.19                           & 66.40\%                           \\ \hline
			VITS-EVC               &  None                     & 4.12                   & 42.92                             & 0.27                   &  99.12\%                        & 4.14±0.08                 & 70.07\%                           \\ 
			EINet (proposed)       & VAD values                                 & \textbf{4.06}  & \textbf{38.28}   & \textbf{0.21} & \textbf{99.48\%} & \textbf{4.38±0.05} & \textbf{75.18\%} \\ \hline
		\end{tabular}%
	}
\end{table*}
As depicted in Table 1, we calculate metrics including mel-cepstral distortion (MCD), root mean squared error of log $F_0$ ($\text{RMSE}_{F_0 }$), average differences of duration (DDUR), and classification accuracy from an external pretrained SER model~\cite{gao2023funasr} ($\text{ACC}_{cls }$) for objective evaluation. In terms of subjective evaluation, mean opinion score (MOS) test is conducted to assess naturalness and emotion similarity of converted audios by 25 participants, each participant assessing 125 utterances in total.

From above-mentioned indicators, it is obvious that the proposed EINet demonstrates competitive performance in both objective and subjective evaluations. 
Notably, in comparison to Seq2seq-EVC, Emovox shows minimal improvement in most metrics, particularly with a reduction of 0.11 in naturalness. This implies that relying solely on direct measurement of distances among acoustic feature for intensity pseudo-labels might neglect inherent emotional variations in a speaker, potentially leading to constrained vocal expression during inference.
In contrast, EINet achieves more realistic intonation and rhythm variations by utilizing VAD values as guidance, resulting in a reduction of $\text{RMSE}_{F_0}$ and DDUR by 4.7 and 0.06, respectively, compared to baseline VITS-EVC. Additionally, there is a significant improvement in naturalness and similarity, which suggests that intensity control module should not compromise basic EVC model when the mismatch between emotion control and speech synthesis is appropriately mitigated. Instead, such controllability has the potential to enhance emotion conversion by capturing more refined emotional cues.
\begin{table}[]
	\centering
	\renewcommand{\arraystretch}{1.1}
	\setlength{\abovecaptionskip}{0.2cm}
	\setlength{\belowcaptionskip}{-0.4cm}
	\caption{Ablation study of proposed method.}
	\label{tab:my-table}
	\resizebox{1.00\columnwidth}{!}{%
		\begin{tabular}{lccc}
			\hline
			\multicolumn{1}{c}{EVC Model}        & $\text{RMSE}_{F_0 }$ $\downarrow$           & DDUR $\downarrow$          & Naturalness $\uparrow$ \\ \hline
			\multicolumn{1}{c}{EINet (proposed)} & \textbf{38.28} & \textbf{0.21} & \textbf{4.38±0.05}    \\
			w/o Identity Maintainer              & 42.65          &  0.23         & 4.29±0.17             \\
			w/o Emotion Renderer                 & 41.42          &  0.29         & 4.18±0.12             \\
			w/o Duration Predictor               & 46.94          &  0.38         & 4.07±0.10             \\ \hline
		\end{tabular}%
	}
\end{table}
\begin{figure}[]
	\centering
	\setlength{\abovecaptionskip}{0.2cm}
	\setlength{\belowcaptionskip}{-0.4cm}
	\scalebox{1}[1.00]{
		\includegraphics[width=\columnwidth]{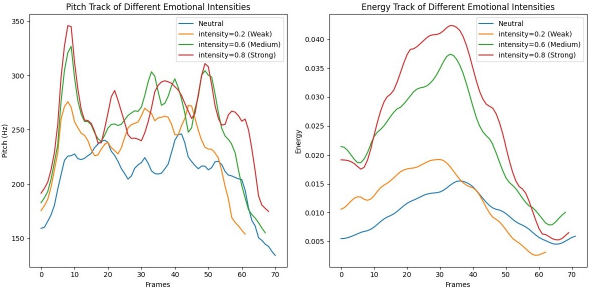}
	}
	\caption{Pitch and energy tracks of a testing clip.}
	\label{Fig2}
\end{figure}

\subsection{Ablation study}
We further conduct an ablation study to validate different contributions. We remove identity maintainer, emotion renderer, and duration predictor in turn and let participants evaluate naturalness of converted audios. From Table 2, we can see that all scores including $\text{RMSE}_{F_0 }$, DDUR, and naturalness are degraded with the removal of different components. 
When remove identity maintainer, a significant increase in $\text{RMSE}_{F_0 }$ is observed. It is attributed that speaker characteristics are not constrained by $L_{F_0 }$ in posterior network, which results in unnatural variations in synthesized intonation.
To further show the contribution of emotion renderer, we replace it with a simple concatenation, resulting in a slight increase in DDUR, which due to the absence of feature fusion between linguistic content and emotional information before monotonic alignment, thereby weakening prior network's modeling of rhythmic changes.
Additionally, the removal of duration predictor leads to a direct impact on all metrics, highlighting the importance of EINet's ability to dynamically adjust speech duration based on the target emotion category and controllable emotional intensity.

\subsection{Controllability of emotional intensity}
To showcase the controllability of emotional intensity, we visualize pitch and energy tracks of voicing parts in testing clips (from neutral to happy), as exemplified in Figure 2. 
It can be observed that as emotional intensity increases, i.e., the induction of emotional states progresses from weak to strong, there is a concurrent broadening of pitch fluctuation and an elevation in peak energy.
Furthermore, Figure 3 presents synthesized Mel-spectrograms with F0 contours, demonstrating that with an increase in emotional intensity, the acoustic variation becomes more pronounced, coupled with more short pauses. This implies that EINet can adaptively  convey intrinsic emotional states based on controllable emotional intensity, achieving optimal outcomes in both intonation and rhythm synthesis.

\subsection{Diversity of emotional samples}
The diversity among emotional samples with varying intensity can be quantified through mean squared distance (MSD) metric, which gauges the pairwise distance distribution of converted audio.
Table 3 elaborates the MSD values for each emotion conversion. Since Emovox did not conduct experiments on \textit{Neu-Sur} in their paper\cite{zhou2022emotion}, the presentation of results is consequently absent here.

It is evident that the proposed EINet achieved optimal outcomes for all transformation pairs. This underscores the effectiveness of utilizing VAD values to accurately capture differences in emotional states, offering a valuable solution for addressing the disparity in emotional intensity modeling and runtime conversion.
Notably, \textit{Neu-Sad} (long duration, lowest VAD values) and \textit{Neu-Sur} (long duration, moderate valence, high arousal and dominance values) outperform others, indicating that the duration predictor is particularly sensitive to duration and VAD values, when modeling rhythmic variations. Consequently, it generates speech expressions with obvious and natural emotional differences, enhancing overall diversity of converted audios.
\begin{table}[]
	\centering
	\renewcommand{\arraystretch}{1.1}
	\setlength{\abovecaptionskip}{0.2cm}
	\setlength{\belowcaptionskip}{-0.4cm}
	\caption{Diversity evaluation of emotional samples.}
	\resizebox{1.00\columnwidth}{!}{%
		\begin{tabular}{c|cccc}
			\hline
			\multirow{2}{*}{EVC Model} & \multicolumn{4}{c}{MSD $\uparrow$}                                           \\ \cline{2-5} 
			& Neu-Ang        & Neu-Hap        & Neu-Sad        & Neu-Sur        \\ \hline
			Emovox                     & 17.87          & 16.88          & 19.86          & -              \\
			EINet (proposed)           & \textbf{19.61} & \textbf{20.55} & \textbf{21.54} & \textbf{22.24} \\ \hline
		\end{tabular}%
	}
\end{table}
\begin{figure}[]
	\centering
	\setlength{\abovecaptionskip}{0.0cm}
	\setlength{\belowcaptionskip}{-0.4cm}
	\scalebox{1}[1.1]{
		\includegraphics[width=\columnwidth]{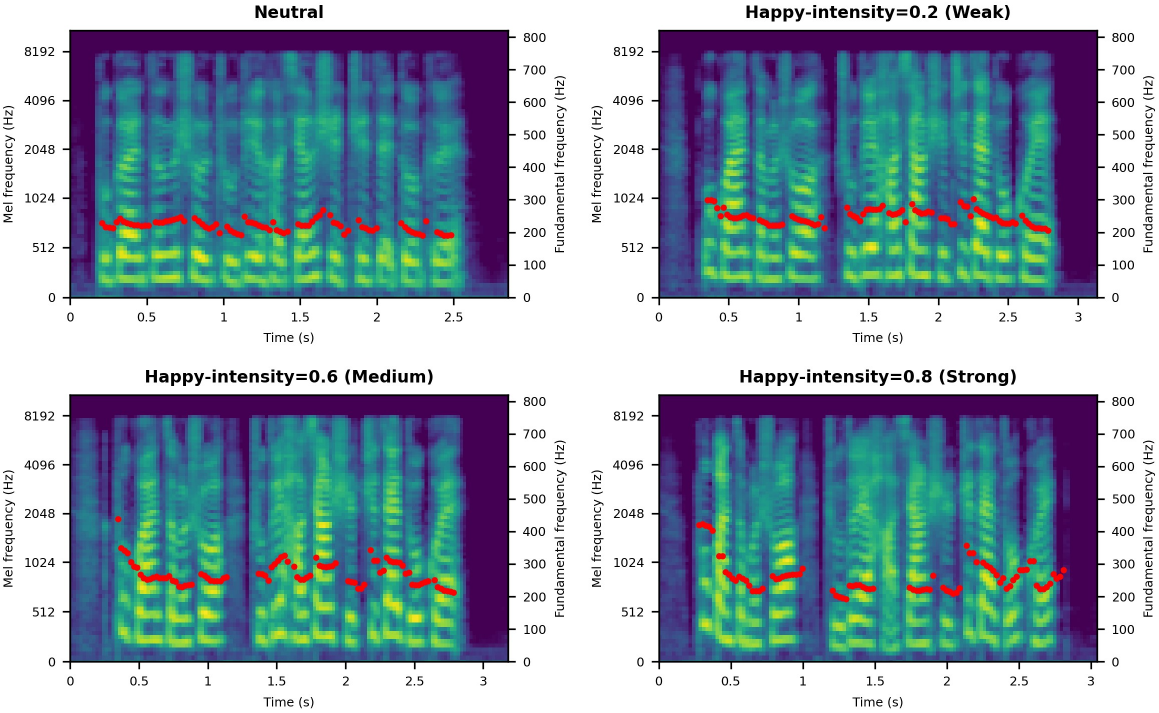}
	}
	\caption{Mel-spectrograms and F0 contours of converted audios at different emotional intensity.}
	\label{Fig3}
\end{figure}

\section{Conclusion}
In this paper, we propose the Emotional Intensity-aware Network (EINet) to achieve realistic emotional voice conversion (EVC) by utilizing controllable emotional intensity. Experimental results on ESD corpus demonstrate its superior performance in enhancing naturalness and diversity of emotional expression, even without explicit emotional intensity annotations. In the future, we will explore the text-based emotion editing for EVC to enhance selectable controllablity of converted audio.

\section{Acknowledgements}
This work was supported in part by the National Key R \& D Project under the Grant 2022YFC2405600, in part by the NSFC under the Grant U2003207 and 61921004, in part by the Jiangsu Frontier Technology Basic Research Project under the Grant BK20192004, and in part by the YESS Program by CAST under the Grant 2023QNRC001.

\bibliographystyle{IEEEtran}
\bibliography{mybib}

\end{document}